\documentclass[times,authoryear,onecolumn,final]{elsarticle}

\usepackage{jasr}
\usepackage{framed,multirow}

\usepackage[table,xcdraw]{xcolor}

\usepackage{colortbl}
\usepackage{array}

\usepackage{amssymb}
\usepackage{latexsym}
\usepackage{subcaption}
\usepackage{soul}

\usepackage[switch]{lineno}

\usepackage{url}
\usepackage{xcolor}
\definecolor{newcolor}{rgb}{.8,.349,.1}
\DeclareUnicodeCharacter{2009}{\,}

\usepackage[citebordercolor=white]{hyperref}

\journal{Advances in Space Research}

\begin{document}

\verso{Ayisha M Ashruf \textit{etal}}

\begin{frontmatter}

\title{Deciphering Solar Cycle Influence on Long-Term Orbital Deterioration of Low-Earth Orbiting Space Debris}%

\author{Ayisha \snm{M Ashruf}\textsuperscript{a,b}\corref{cor1}}
\cortext[cor1]{Corresponding author: }
\ead{ayisha@vssc.gov.in, ayishamashruf@gmail.com}
\author{Ankush \snm{Bhaskar}\textsuperscript{a}}
\author{C. \snm{Vineeth}\textsuperscript{a}}
\author{Tarun \snm{Kumar Pant}\textsuperscript{a}}

\affiliation[1]{organization={Space Physics Laboratory},
	addressline={Vikram Sarabhai Space Centre},
	city={Thiruvananthapuram, Kerala},
	postcode={695022},
	country={India}}
\affiliation[2]{organization={Indian Institute of Space Science and Technology},
	addressline={Valiamala},
	city={Thiruvananthapuram, Kerala},
	postcode={695547},
	country={India}}


\begin{abstract}

The rapid increase in the number of space debris represents a substantial threat to the sustained viability of space operations and underscores the importance of understanding long-term drivers of orbital decay. This first of its kind study examines the long-term impact of solar activity on the orbital decay of 17 LEO debris objects across Solar Cycles 22, 23, and 24 using Two-Line Element (TLE) data spanning these three cycles. Analysis of TLE-derived decay profiles, in conjunction with sunspot numbers (SSN) and F10.7 index, reveals a threshold: orbital decay rates increase sharply when SSN exceeds approximately 67–75\% of its cycle peak. This threshold corresponds to enhanced thermospheric density driven by elevated solar input, resulting in increased atmospheric drag. The orbital decay rates at the peak of each solar cycle show a progressive decline from Cycle 22 to Cycle 24, mirroring the corresponding decrease in solar activity. Decay profiles for Solar Cycle 24, predicted using ballistic coefficients derived from TLE data during Cycles 22 and 23 and atmospheric densities from the MSIS 2.0 model, show strong agreement with observations after applying a scaling factor. However, two high-inclination objects exhibited significant deviations, highlighting limitations in the MSIS 2.0 model's ability to represent atmospheric conditions at high latitudes. In contrast, lower-inclination objects showed excellent correspondence. Overall, the findings confirm solar-driven thermospheric variability as the dominant factor influencing long-term orbital decay and emphasize the need to refine atmospheric models—particularly for polar regions—to improve re-entry predictions and satellite mission planning.

\end{abstract}

\begin{keyword}
\KWD space debris\sep solar cycle\sep orbital decay\sep LEO
\end{keyword}

\end{frontmatter}


\section{Introduction}
\label{sec1}

Space debris - ranging from defunct satellites and discarded rocket stages to fragments from collisions - poses an ever‐increasing threat to active spacecraft and human spaceflight \citep{GRAHAM1999, CHUNLAI2002,KLEINIG2022,SMITH2020,HAYES2023}. As the space industry continues to expand, so does the number of objects in orbit, presenting considerable hazards to functioning satellites, ongoing space missions, and astronaut safety. A pressing issue arises from the risk of cascading collisions, commonly termed the Kessler Syndrome \citep{kessler2010kessler}, where a single collision can generate substantial additional debris, triggering a domino effect of subsequent collisions. While numerous studies have investigated collision risks and mitigation strategies, less attention has been paid to the long‐term physical processes that drive the orbital decay of debris.

One of the key factors influencing orbital decay is the solar activity \citep{nwankwo2018space,klinkrad2006space, walterscheid1989solar}. Typically, every 11 years, the Sun undergoes active and quiet phases, constituting a solar cycle, resulting in notable variations in the emitted flux of electromagnetic and corpuscular radiation \citep{lean1987solar, hathaway2015solar, richardson2012near, gopalswamy2006properties}. Increased solar emissions during active phases heat and expand the thermosphere, thereby raising atmospheric density at orbital altitudes and enhancing drag experienced by orbiting bodies \citep{harris1962theoretical, CLETTE2015, weng2020machine, CRISP2021}. This makes the solar cycle variations a crucial component in long-term orbital dynamics. Recent events such as the loss of Starlink satellites linked to mild geomagnetic disturbances \citep{Dang2022,Kataoka2022} further underscore the operational importance of understanding these effects.

While the influence of solar activity on satellite drag is well recognized, a systematic investigation into its long-term impact on the orbital decay of space debris remains lacking. To date, there has been no such study spanning multiple solar cycles. For instance, \cite{nwankwo2018space} discussed the typical orbital decay of a LEO satellite between 1999 and 2014, clearly showing a correlation between solar activity and orbital decay. However, this study covered only a 15-year period. In the present work, we address this gap by analyzing the orbital decay of 17 LEO debris objects—filtered from an initial set of 95 by excluding medium Earth orbit (MEO) debris, objects with higher eccentricity orbits, and objects above 800 km in altitude. Using TLE data spanning three consecutive solar cycles (Solar Cycles 22, 23, and 24), we correlate decay rates with solar activity, utilizing over 36 years of sunspot numbers and F10.7 solar flux data. We further attempt to model the orbital decay profiles of these objects by estimating their ballistic coefficients from historical TLE data and using atmospheric density values from the NRLMSIS 2.0 model \citep{Emmert2021}. The methodology is outlined in the next section. This analysis not only enhances our understanding of debris dynamics but also provides valuable insights for improving space situational awareness, collision avoidance, and debris mitigation strategies.

The novelty of this research lies in its long-term, data-driven approach that bridges the gap between short-term space weather events and sustained orbital decay trends. Through a comprehensive analysis of historical TLE data, this study offers valuable insights into the behavior of space debris under varying solar conditions across multiple solar cycles, thereby contributing to a better understanding of orbital decay processes and more sustainable space operations.

\section{Data and Methodology}

This study investigates the long-term impact of solar activity on the orbital decay of space debris. The sunspot number constitutes a historical time series spanning from 1700 to the present, capturing the 11-year cyclic and long-term variations in solar activity and, consequently, space weather. Although contemporary observations offer more refined parameters for understanding space-weather impacts, the SSN represents the earliest direct record of solar activity. It serves as an essential link connecting past and present solar behavior \cite{CLETTE2014, CLETTE2016,Jayalekshmi2022}. Along with the SSN data, the F10.7 index is also known to be a fairly good representative of solar activity \cite{chen2011does}. The study covers three full solar cycles: Solar Cycle 22 (September 1986 – August 1996), Solar Cycle 23 (September 1996 – December 2008), and Solar Cycle 24 (January 2009 – December 2019). Data for SSN and F10.7 solar radio flux index were obtained from the GFZ Potsdam database (\url{www.gfz-potsdam.de})

The orbital parameters of satellites and space debris are commonly represented using Two-Line Elements, which consist of up to 69 alphanumeric characters. A detailed description of the TLE format can be found in \cite{VALLADO2012}. For this study, space debris originating in the 1960s were selected, having continuous TLE data with sufficient temporal resolution across Solar Cycles 22, 23, and 24. Importantly, all selected objects remained in orbit as of June 2023. An initial set of 95 debris objects was identified using Space-Track website (\url{www.space-track.org}). To refine the dataset for consistent orbital characteristics, we excluded debris in MEO, those with high eccentricities, and those at altitudes above 800 km. This filtering resulted in a final dataset of 17 LEO debris objects, as listed in Table \ref{tab:Tab1}. The corresponding TLE data, covering the period from 01 September 1986 to 30 June 2023, were obtained from the Space-Track website and used for the analysis. Given that the objects analyzed in this study are located below 800 km in altitude, the influence of solar radiation pressure is minimal and was thus excluded \citep{SANG2013}. Additional physical characteristics — mass and maximum, minimum, and average cross-sectional areas — were obtained from the DISCOS (Database and Information System Characterising Objects in Space) database (\url{https://discosweb.esoc.esa.int/}) where available, and are also included in Table \ref{tab:Tab1}.

\begin{table}[]
	\centering
	\caption{Satellite Catalog (SATCAT) Number, satellite names, orbital inclinations (i), masses, maximum, minimum and average cross-sectional areas of the 17 LEO objects}
	\begin{tabular}{c|c|c|c|ccc}
		SATCAT No. & SATNAME            & i ($^{\circ}$) & mass (kg) & A$_{max}$ (m$^{2}$) & A$_{min}$ (m$^{2}$) & A$_{avg}$ (m$^{2}$) \\ \hline
		22       & EXPLORER 7         & 50.28   & 41.13     & 0.454        & 0.454        & 0.454        \\
		29       & TIROS 1            & 48.38   & 118.93    & 1.036        & 0.514        & 0.853        \\
		45       & TRANSIT 2A         & 66.7    & 100.1     & 0.65         & 0.65         & 0.65         \\
		46       & SOLRAD 1 (GREB)    & 66.69   & 18.83     & 0.204        & 0.204        & 0.204        \\
		115      & THOR ABLE DEB (YO) & 48.16   & -         & -            & -            & -            \\
		162      & TIROS 3            & 47.9    & 127.85    & 1.036        & 0.514        & 0.853        \\
		227      & DELTA 1 DEB (YO)   & 48.15   & -         & -            & -            & -            \\
		228      & DELTA 1 DEB (YO)   & 48.43   & -         & -            & -            & -            \\
		262      & THOR ABLESTAR DEB  & 66.46   & -         & -            & -            & -            \\
		309      & TIROS 5            & 58.09   & 127.85    & 1.081        & 0.599        & 0.92         \\
		397      & TIROS 6            & 58.31   & 125.87    & 1.081        & 0.599        & 0.82         \\
		716      & TIROS 8            & 58.5    & 117.94    & 1.081        & 0.599        & 0.92         \\
		720      & DELTA 1 DEB        & 58.48   & 1         & 12.25        & 1            & 6.125        \\
		733      & THOR AGENA D R/B   & 99.04   & 600       & 9.614        & 1.767        & 8.306        \\
		734      & OPS 3367 A         & 99.01   & 128.84    & 0.412        & 0.283        & 0.377        \\
		876      & COSMOS 44          & 65.06   & 1250      & 6.485        & 1.539        & 5.718        \\
		877      & SL-3 R/B           & 65.08   & 1427.16   & 11.216       & 5.309        & 10.414      
	\end{tabular}
	\label{tab:Tab1}
\end{table}

The TLE elements for each debris object were used to construct a database of orbital parameters across different epochs. This database served as the foundation for all subsequent analyses. The semi-major axis '$a$' of the orbit of each space debris at a given epoch is obtained from mean motion using the following equation, 

\begin{eqnarray}
	a = \left(\frac{GM}{\left(\frac{2\pi n}{86400}\right)^{2}}\right)^{\frac{1}{3}}
\end{eqnarray}

where $M$ is the mass of the Earth, $G$ is the universal gravitational constant and $n$ is the mean motion in revolutions per day.

Once the time series of semi-major axes was established, the altitudes were computed and plotted alongside sunspot number as a function of time. To handle outliers within the data points, we applied a z-score filtering method: any data point with a semi-major axis z-score greater than 3 was considered an outlier and excluded from the analysis.

To better quantify the impact of solar activity, we calculated the decay rates for all 17 objects during the peaks of Solar Cycles 22, 23, and 24. The resulting values, for each cycle, are listed in Table \ref{tab:Tab2}. Additionally, we examined the relationship between the initial altitude of the debris and its decay rate near the peak of each solar cycle. Histograms were generated to visualize the distribution of peak decay rates across the different cycles. Furthermore, we examined the correlation between orbital decay rates and SSN and the F10.7 index.

The method proposed by \cite{SANG2013} was used to estimate the ballistic coefficient (BC) of LEO space debris based on historical TLE data. The BC is computed as $C_{d}A/m$ where $C_d$ is the drag coefficient, $A$ is the cross-sectional area and $m$ is the mass of the object. In this approach, BC is derived from variations in the mean semi-major axis extracted from the TLEs. Data from Solar Cycles 22 and 23 were utilized for BC estimation, using the following equation:

\begin{eqnarray}
	BC = -\frac{\mu \Delta^{t_2}_{t_1}}{\sum_{t = t_1}^{t_2}a^2 \rho v. v_{rel}^2. \Delta t}
\end{eqnarray}

where $\mu$ is the product of the universal gravitational constant and Earth's mass, $a$ is the semi-major axis of the debris, $\rho$ is the atmospheric density, $v$ is the satellite velocity, and $v_{rel}$ is the velocity of the debris relative to the co-rotating atmosphere.  Since the atmospheric velocity is significantly lower than the satellite velocity — with the average rotational speed of the atmosphere at 400 km being about 180 m/s \citep{KINGHELE1970}, compared to satellite velocities of around 7 km/s — we approximate $v_{rel}$ as $v$. The estimated BC values were subsequently used to predict the orbital decay of the debris objects during Solar Cycle 24, and the results were compared against the actual decay profiles derived from TLE data for the same period.

To model the decay profile, initial TLE data at the beginning of Solar Cycle 24 is utilized to propagate the orbit every 15 seconds until the end of Solar Cycle 24, in order to obtain the latitude-longitude values. The altitude is determined using the drag equation to estimate changes in orbital velocity. The change in velocity resulting from atmospheric drag over a time interval, $\Delta t$, can be expressed as:

\begin{eqnarray}
	\Delta v = \frac{1}{2} \rho v^2 C_D \frac{A}{m} \Delta t
\end{eqnarray}

For a satellite in orbit around the Earth, the gravitational force is balanced by the centripetal force.

\begin{eqnarray}
	\frac{GMm}{(R_E + h)^2} = \frac{mv^2}{R_E + h}
\end{eqnarray}

where, $G$ is the universal gravitational coefficient, $M$ is the mass of the Earth, $R_{E}$ is the radius of the Earth, $h$ is the altitude of the satellite and m is the satellite mass. This equation can be re-arranged to obtain the equation for the altitude,

\begin{eqnarray}
	h = \frac{GM}{v^2} - R_E
\end{eqnarray}

The resulting new altitude due to the drag force can be computed as,

\begin{eqnarray}
	h^{'} = \frac{GM}{(v + \Delta v)^2} - R_E
\end{eqnarray}

The drag coefficient quantifies the interaction between the surface of the spacecraft and the impinging atmospheric molecules within the free-molecular flow regime \citep{DOORNBOS2006}. Although the drag coefficients of these objects are not directly available, the computed BC values were used to estimate the drag coefficients for 13 objects with known mass and cross-sectional area information.

\section{Observations}

\begin{figure}[ht]
	\centering
	\includegraphics[width=0.8\textwidth]{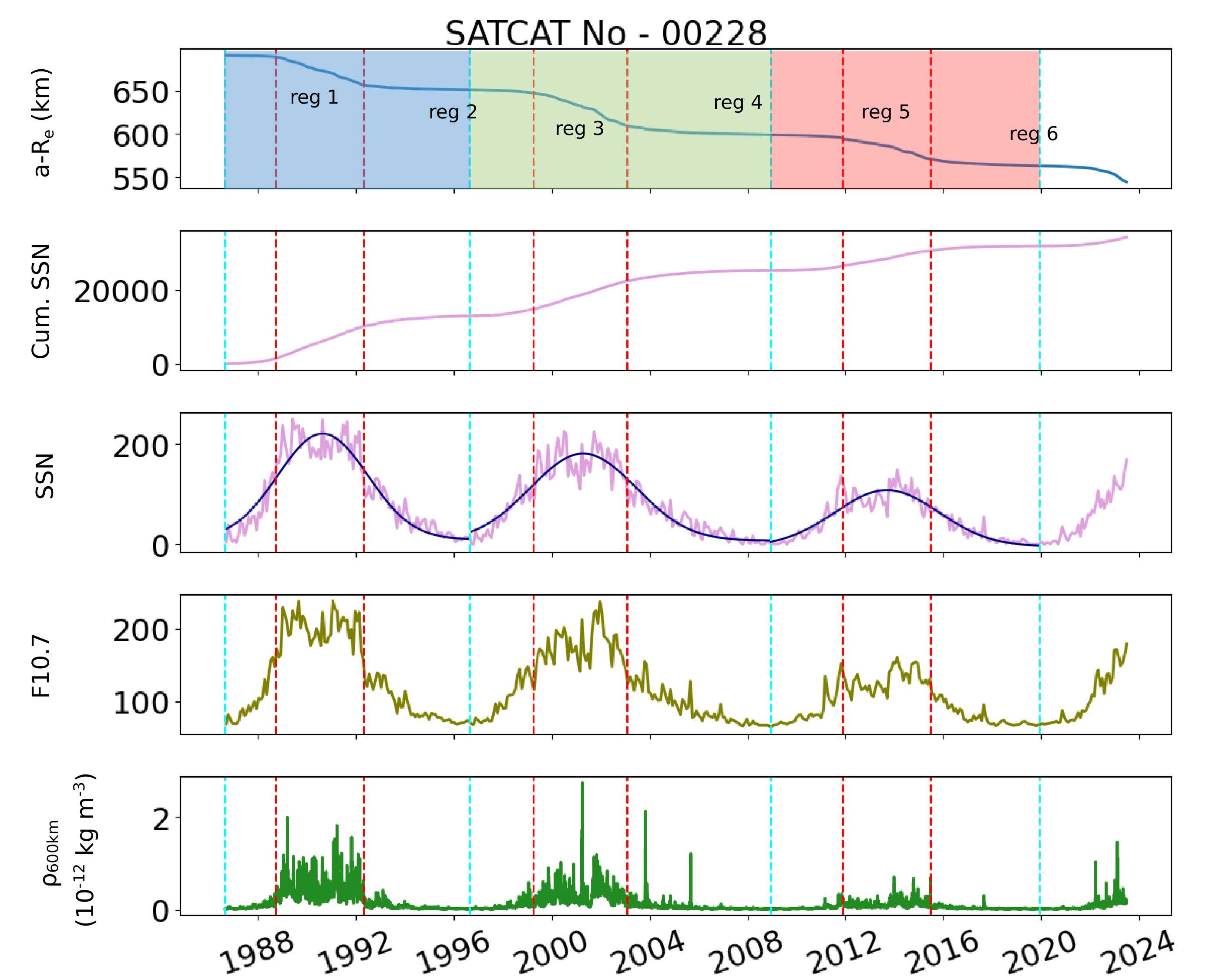}
	\caption{Variation of apparent altitude, cumulative sunspot number, sunspot number, F10.7 (adjusted, s.f.u.), and MSIS model-derived mean global density at 600 km over three solar cycles for a sample debris object}
	\label{fig:Fig1}
\end{figure}

Figure \ref{fig:Fig1} shows the orbital decay of a representative LEO debris object (Delta 1 debris) alongside solar activity indicators — SSN and F10.7 index — over Solar Cycles 22, 23, and 24. In the top panel, the apparent altitude ($a - R_e$) — obtained by subtracting Earth's mean radius from the mean semi-major axis derived from TLE data — is plotted. The object’s altitude decreases from roughly 690 km at the start of Solar Cycle 22 to about 550 km at the end of Solar Cycle 24. The three solar cycles are indicated by color-coded blocks (blue for Cycle 22, green for Cycle 23, and red for Cycle 24). The vertical red dashed lines serve as markers for areas characterized by nearly linear and rapid orbit decay during each solar cycle. To position these red lines, the rate of decay is calculated for the entire dataset using two adjacent points. The points where a sharp change in the decay rate occurs is identified as the starting and ending points for slope calculations.


The second panel depicts the cumulative sunspot number, which rapidly increases during solar maximum and levels off during the minimum. The third panel presents the SN, along with a fitted Gaussian curve (shown in blue). In Cycle 22, the red dashed lines intersect the Gaussian-fitted sunspot curve at approximately 63\% and 71\% of its peak during the ascending and descending phases, respectively. For Cycle 23, these intersections occur at about 68\% and 75\%, while for Cycle 24 the threshold is around 67\% for both phases. This consistency suggests that when the SSN exceeds roughly 67\% of its maximum value, a transition to rapid orbital decay is observed. The F10.7 index values are plotted in the fourth panel, and the MSIS 2.0 model-derived global mean atmospheric density at 600 km is plotted in the bottom panel. From Figure \ref{fig:Fig1}, it can be seen that Solar Cycle 22 exhibits the highest overall solar activity, followed by Cycle 23, while Cycle 24 is comparatively weak—a pattern mirrored by the varying steepness of the orbital decay slopes. As expected, the decay rate is significantly higher during solar maxima and becomes more gradual as the cycle approaches its minimum. The consistent intersection of the red dashed lines with the Gaussian-fitted sunspot number curves at around 67–75\% of their peak values, suggests that once solar activity surpasses this threshold, increased thermospheric density drives a rapid increase in orbital decay. The overall behaviour is consistent with the established understanding that heightened solar activity increases the thermospheric scale height and density, primarily due to enhanced X-ray and extreme ultraviolet (EUV) emissions from the Sun \citep{Mlynczak2018}. As the drag force experienced by orbiting objects is strongly influenced by atmospheric density, this leads to a corresponding increase in orbital decay. The acceleration experienced by the object due to drag force is expressed as,

\begin{eqnarray}
	a_{drag} = -\frac{1}{2}\frac{C_{D}A}{m}\rho v_{rel}^{2}
\end{eqnarray}

where $m$ is the mass of the orbiting body, $C_{D}$ is the drag coefficient, $A$ is the cross-sectional area, $\rho$ is the atmospheric density and $v_{rel}$ is the relative velocity of the body with respect to the atmosphere.

\begin{figure}[hbt!]
	\centering
	\includegraphics[width=0.7\textwidth]{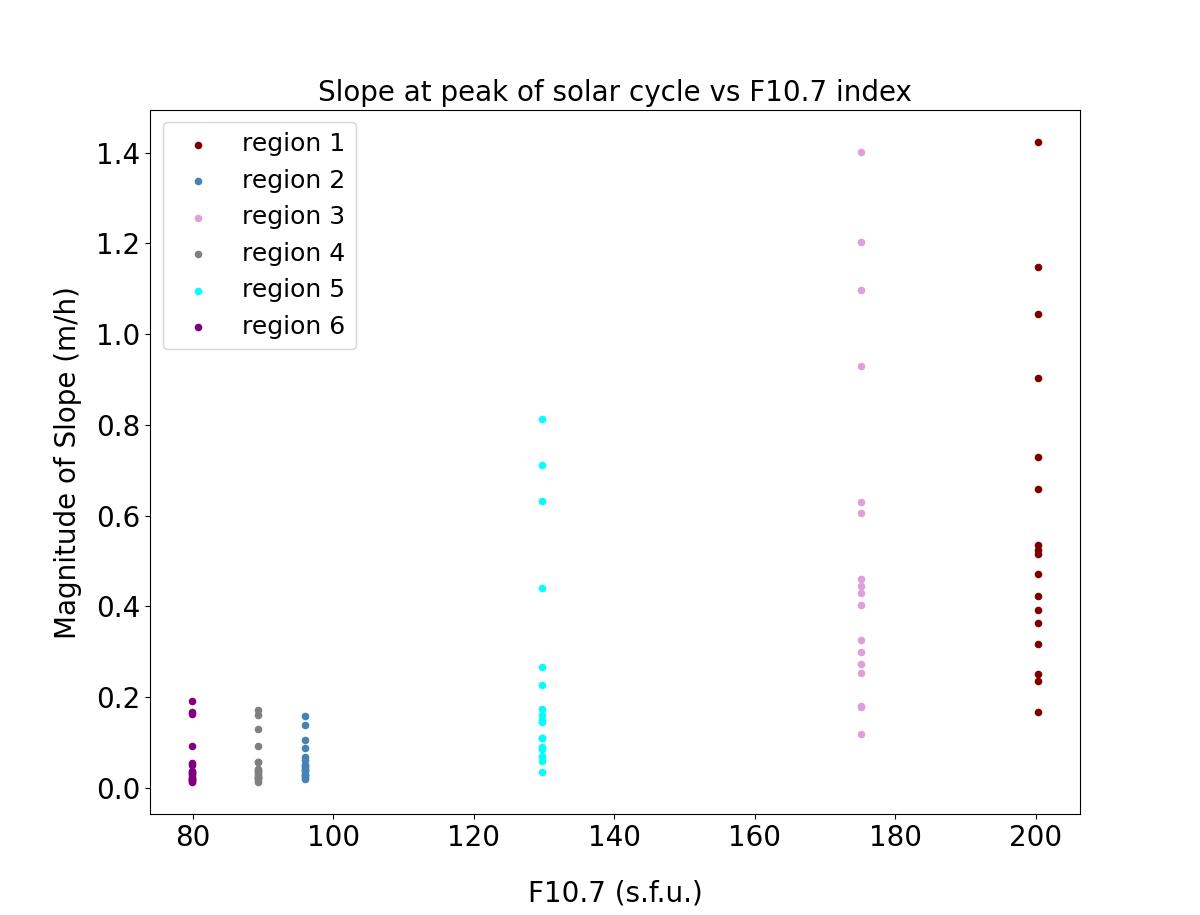}
	\caption{Relationship between F10.7 solar flux and decay rate}
	\label{fig:Fig2}
\end{figure}

In Figure \ref{fig:Fig1}, six distinct regions can be identified, each characterized by a different rate of orbital decay. Three of these regions — marked by red dashed lines corresponding to the peaks of the solar cycles — display noticeably steeper slopes, indicating periods of enhanced decay, while the remaining three show a more gradual decline. Figure \ref{fig:Fig2} depicts the slopes computed over these six segments, alongside the average F10.7 index during each interval. As expected, regions with higher F10.7 values (regions 1, 3, and 5) correspond to steeper slopes, reflecting the higher solar input to the Earth's atmosphere during those periods.

\begin{figure}[hbt!]
	\centering
	\includegraphics[width=0.8\textwidth]{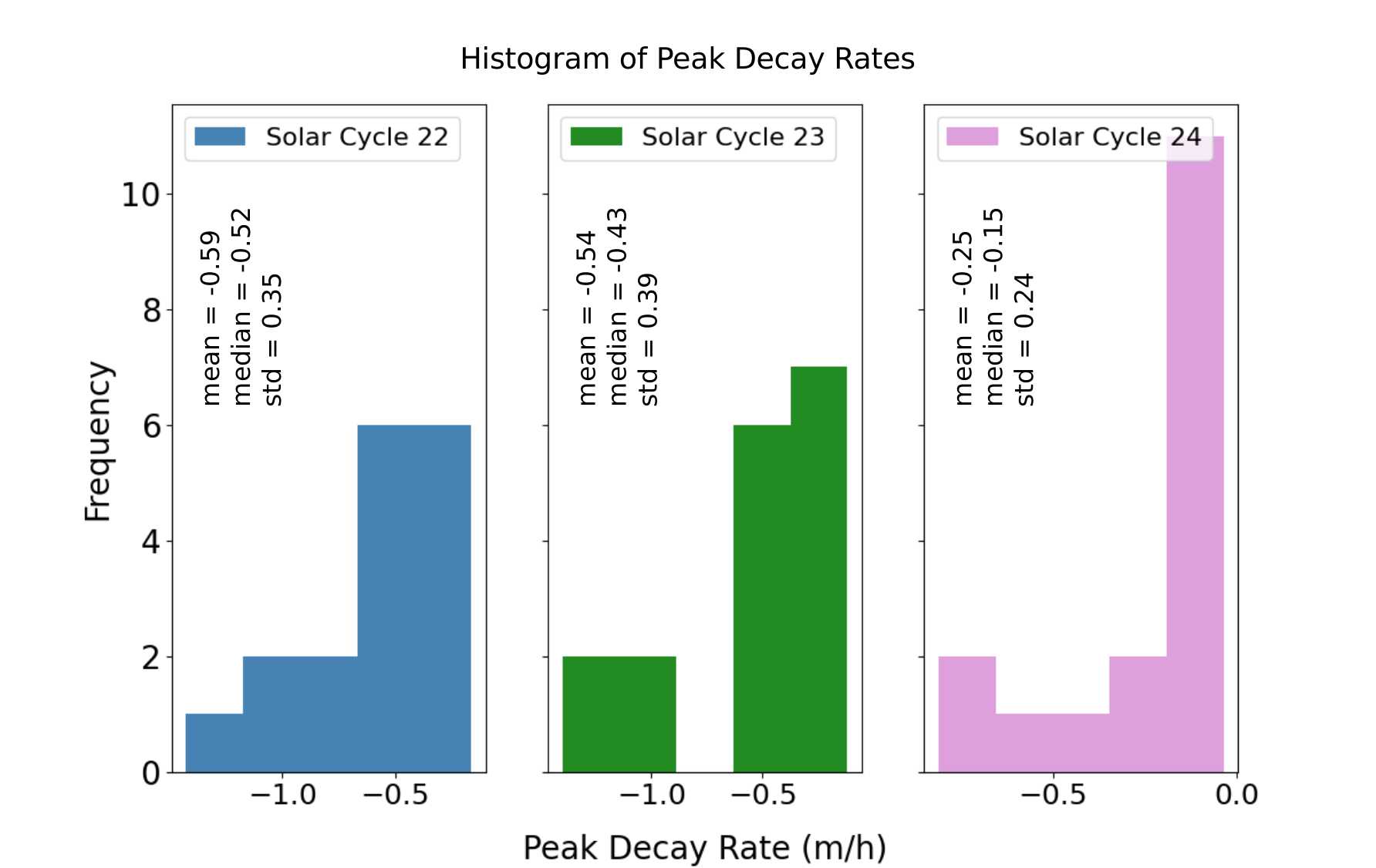}
	\caption{Distribution of peak decay rates during solar cycles 22, 23 and 24}
	\label{fig:Fig3}
\end{figure}

Figure \ref{fig:Fig3} presents histograms of the peak orbital decay rates—estimated as the slopes of the curves enclosed within the red vertical lines in Figure \ref{fig:Fig1}—for 17 debris objects across Solar Cycles 22, 23, and 24. Each cycle reveals distinct patterns that reflect the varying levels of solar activity. Solar Cycle 22 exhibits the highest decay rates, with a mean of –0.59 m/h and a median of –0.52 m/h. Most objects during this cycle experience relatively steeper decay, consistent with strong solar activity that enhances thermospheric density and, consequently, atmospheric drag. Solar Cycle 23 shows slightly lower decay rates, with a mean of –0.54 m/h and a median of –0.43 m/h. In contrast, Solar Cycle 24 displays a marked reduction in decay rates. The mean drops to –0.25 m/h and the median to –0.15 m/h. The decay rates are more tightly clustered around zero, reflecting weak solar activity and reduced thermospheric heating—resulting in lower atmospheric drag and slower orbital decay. Overall, the magnitude of decay rates decreases progressively from Cycle 22 to Cycle 24, closely mirroring the corresponding decline in solar activity. These histograms strongly supports the connection between solar activity and orbital decay, showing that peak decay rates are directly influenced by the intensity of solar cycles. Higher solar input, indicated by increased F10.7 flux and sunspot numbers, leads to enhanced thermospheric density, which in turn accelerates orbital decay.

\begin{table}[]
	\centering
	\caption{Decay rates (ROD) during the peak of solar cycles 22, 23 and 24 for the 17 LEO debris objects.}
	\begin{tabular}{c|c|c|c}
		SATCAT No. & SC 22 (m/h) & SC 23 (m/h) & SC 24 (m/h) \\ \hline
	22       & -1.42                & -1.40               & -0.81                \\
	29       & -0.52                & -0.46               & -0.17                \\
	45       & -0.39                & -0.30               & -0.11                \\
	46       & -0.73                & -0.61               & -0.23                \\
	115      & -0.90                & -0.93               & -0.44                \\
	162      & -0.24                & -0.18               & -0.06                \\
	227      & -0.42                & -0.33               & -0.11                \\
	228      & -1.04                & -1.20               & -0.71                \\
	262      & -1.15                & -1.10               & -0.63                \\
	309      & -0.53                & -0.45               & -0.16                \\
	397      & -0.66                & -0.63               & -0.27                \\
	716      & -0.47                & -0.40               & -0.15                \\
	720      & -0.52                & -0.43               & -0.15                \\
	733      & -0.25                & -0.18               & -0.07                \\
	734      & -0.17                & -0.12               & -0.04                \\
	876      & -0.36                & -0.27               & -0.09                \\
	877      & -0.32                & -0.25               & -0.09          
	\end{tabular}
	\label{tab:Tab2}
\end{table}

The ballistic coefficients for the 17 debris objects were initially estimated using TLE data from Solar Cycles 22 and 23, combined with atmospheric densities from the MSIS 2.0 model at the corresponding locations. The MSIS 2.0 model requires latitude, longitude, altitude, and epoch as input parameters, with the latter three obtained directly from the TLEs. Latitude and longitude were derived using the EarthSatellite module from the Skyfield Python library. These initial BC estimates were then used to predict the altitude decay profiles of 17 objects during Solar Cycle 24. However, it was observed that the predicted profiles did not align with the TLE-derived profiles unless the BCs were scaled by a factor $k$. The derived BCs, the applied scaling factors, and the resulting scaled BCs are summarized in Table \ref{tab:Tab3}.

\begin{figure}[hbt!]
	\centering
	\includegraphics[width=0.7\textwidth]{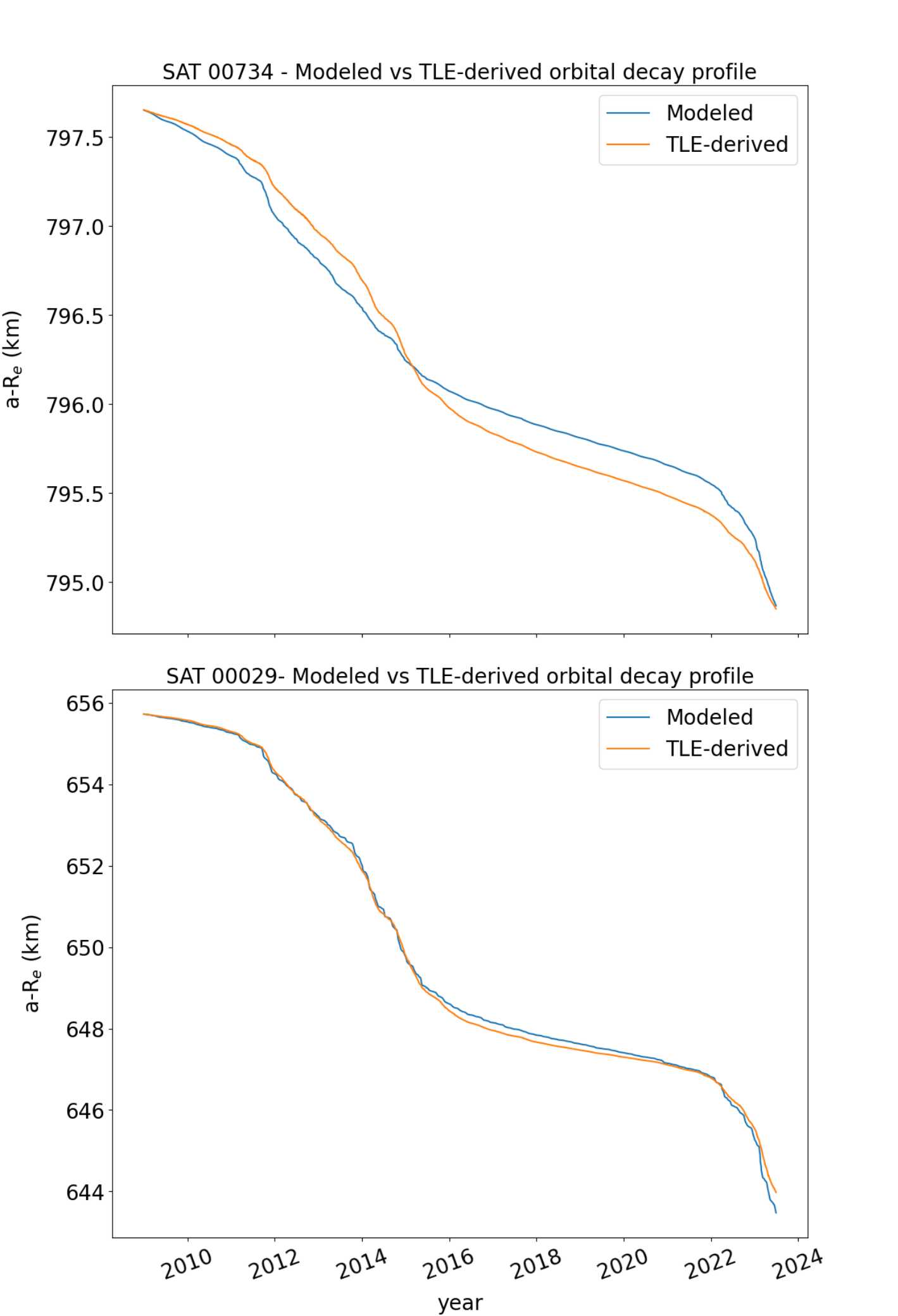}
	\caption{TLE-derived and modeled altitude decay profiles of (top panel) SAT 734 (bottom panel) SAT 029}
	\label{fig:Fig4}
\end{figure}

Interestingly, for two debris objects—SAT 733 and SAT 734—the decay profiles could not be accurately reproduced, even after applying various scaling factors. Unlike the other objects, which had inclinations between 45$^\circ$ and 70$^\circ$, these two were in highly polar orbits with inclinations near 99$^\circ$. The top panel of Figure \ref{fig:Fig4} shows the best-fit modeled decay for SAT 734. Although both the modeled and TLE-derived profiles exhibit a general downward trend consistent with orbital decay, significant discrepancies exist, especially over extended time periods. This suggests potential limitations in the MSIS 2.0 model’s ability to accurately represent atmospheric densities in polar regions. In contrast, the bottom panel of Figure \ref{fig:Fig4} presents the decay profile for SAT 029, where the modeled and TLE-derived curves show excellent agreement. As shown in Table \ref{tab:Tab3}, for the 15 objects with well-matched profiles (excluding SAT 733 and SAT 734), the scaling factors ranged from 0.55 to 0.79, with a mean value of 0.71.

The scaled TLEs have been used to calculate the C$_{d}$ values for objects where both mass and cross-sectional area information are available, which applies to 13 of the objects. These values are presented in Table \ref{tab:Tab3}. However, since the predicted decay profiles for objects 733 and 734 do not closely match the actual TLE-derived curves, the BC values—and consequently the C$_{d}$ values—derived for these two objects can be disregarded.

\begin{table}[]
	\centering
	\caption{Derived BCs, scaling factors (k), scaled BCs, and the minimum, maximum, and average drag coefficients for the 17 LEO debris objects.}
	\begin{tabular}{c|c|c|c|ccc}
		SATCAT No. & BC$_{derived}$ & k    & BC$_{scaled}$ & C$_d$ (min) & C$_d$ (max) & C$_d$(avg) \\ \hline
		22       & 0.0598     & 0.64 & 0.0383    & 3.466     & 3.466     & 3.466    \\
		29       & 0.0193     & 0.74 & 0.0142    & 1.636     & 3.297     & 1.987    \\
		45       & 0.0533     & 0.71 & 0.0378    & 5.826     & 5.826     & 5.826    \\
		46       & 0.0657     & 0.79 & 0.0519    & 4.790     & 4.790     & 4.790    \\
		115      & 0.0383     & 0.74 & 0.0284    & -       & -       & -      \\
		162      & 0.0232     & 0.67 & 0.0155    & 1.917     & 3.864     & 2.328    \\
		227      & 0.0503     & 0.72 & 0.0362    & -       & -       & -      \\
		228      & 0.0304     & 0.78 & 0.0237    & -       & -       & -      \\
		262      & 0.2539     & 0.76 & 0.1930    & -       & -       & -      \\
		309      & 0.0388     & 0.69 & 0.0268    & 3.168     & 5.717     & 3.722    \\
		397      & 0.0177     & 0.73 & 0.0130    & 1.508     & 2.722     & 1.988    \\
		716      & 0.0199     & 0.75 & 0.0150    & 1.632     & 2.945     & 1.917    \\
		720      & 0.0493     & 0.74 & 0.0365    & 0.003     & 0.036     & 0.006    \\
		733      & 0.0343     & 0.85 & 0.0292    & 1.822     & 9.911     & 2.109    \\
		734      & 0.0209     & 0.94 & 0.0197    & 6.157     & 8.964     & 6.729    \\
		876      & 0.0196     & 0.55 & 0.0108    & 2.077     & 8.753     & 2.356    \\
		877      & 0.0182     & 0.69 & 0.0125    & 1.594     & 3.367     & 1.717   
	\end{tabular}
	\label{tab:Tab3}
\end{table}

\section{Discussion}

The rapid expansion of the space sector and the exponential increase in satellites and associated debris have made the monitoring and mitigation of space debris critically important. Many debris objects remain in orbit for years to decades, making the impact of long-term solar activity on their orbital evolution a highly relevant subject for investigation. This study offers a systematic investigation into how solar cycles influence the orbital decay of space debris using 17 LEO debris objects that have been in orbit since 1967. A key finding is the clear correlation between orbital decay of space debris and solar activity, reinforcing the role of solar-driven thermospheric dynamics in shaping orbital lifetimes. 

Notably, studies have shown that satellites initially deployed at approximately 500 km altitude can exhibit lifetimes of up to 30 years during solar minimum conditions, but only about 3 years during solar maximum \citep{walterscheid1989solar}. The study explains that the dominant driver of increased atmospheric density during solar maxima is the rise in thermospheric temperature, primarily due to increased absorption of EUV radiation by atomic oxygen. This increase in temperature also leads to changes in atmospheric composition, especially an increase in the proportion of heavier species like argon. The study further shows that although geomagnetic storms can cause short-lived increases in thermospheric density, these variations are transient and secondary compared to the long-term effects of solar EUV variations over a solar cycle. Similarly, \cite{nwankwo2018space} report that a satellite at 450 km can experience orbital decay rates of up to 41 km per year during solar maxima and around 11 km per year during solar minima. 
The debris objects considered in our analysis are located in the altitude range of 600 to 800 km. During solar maximum conditions for Solar Cycles 22 and 23, these objects experienced decay rates ranging from 1 to 16 km per year. For Solar Cycle 24, decay rates fell between 0.5 and 9 km per year. During solar minimum conditions, for all three cycles, the debris objects experienced orbital decay rates ranging from 100 meters per year to 1 km per year. The variation in decay rates is due to differences in the mass and orientation of the 17 objects. Furthermore, the atmospheric density at altitudes between 600 and 800 km is much lower than at 450 km, which results in reduced drag force and consequently lower overall decay rates.

Interestingly, our analysis reveals a consistent threshold in solar activity—occurring when sunspot numbers exceed approximately 67–75\% of their peak—beyond which orbital decay rates increase rapidly. This suggests the presence of a critical solar input level, above which thermospheric density rises sufficiently to increase atmospheric drag. This threshold has practical implications for predicting re-entry of space debris and improving orbital decay forecasts in response to solar activity.

The variation in decay slopes across Solar Cycles 22, 23, and 24 reflects the relative strength of each cycle: the steeper slopes observed during Cycle 22 align with its higher solar activity, while the more gradual decay during Cycle 24 corresponds to its weaker solar forcing. These observations are consistent with our understanding that increased EUV and X-ray flux during solar maxima enhances thermospheric heating and expansion, resulting in higher atmospheric density and drag at LEO altitudes \citep{Mlynczak2018}.

Furthermore, the strong agreement between elevated decay rates and higher F10.7 values (Figure \ref{fig:Fig2}) supports the conclusion that solar-induced thermospheric variability is the dominant driver of orbital decay. This is further substantiated by the histogram of peak decay rates (Figure \ref{fig:Fig3}), which reveals a systematic decline in both mean and median values from Cycle 22 to Cycle 24.  The tighter clustering of decay rates near zero during Cycle 24 corresponds to significantly lower atmospheric densities associated with reduced solar activity compared to the previous two cycles.

The need for scaling ballistic coefficients to align predicted and observed decay profiles could arise from two primary factors: the method of deriving BCs from historical TLEs is not entirely robust, and the method itself relies on densities provided by atmospheric models such as the MSIS 2.0 model, which can be in error up to 25\%. To accurately predict the orbital decay of the debris objects, the BCs needed to be scaled by values ranging from 0.55 to 0.79. Although the scaling approach proved successful for most of the debris objects, the cases of SAT 733 and SAT 734 highlight significant limitations in the MSIS 2.0 model when applied to objects in highly polar orbits. These two objects, with near-99$^{\circ}$ inclinations, showed considerable discrepancies between the predicted and TLE-derived decay profiles. This suggests that the MSIS 2.0 model may not accurately capture atmospheric conditions in high-latitude regions, where atmospheric densities can vary in ways that the model does not account for. 

In contrast, objects like SAT 029, which follow lower-inclination orbits, showed excellent agreement between the predicted and observed decay profiles. This indicates that the MSIS 2.0 model is more accurate within $\pm$70$^{\circ}$ latitudes. To improve the accuracy of orbital decay predictions for space debris, further refinement of the MSIS 2.0 model is needed, particularly to better capture thermospheric conditions at high latitudes.

\section{Conclusion}

This study provides a comprehensive analysis of the influence of solar activity on the orbital decay of 17 historic LEO space debris objects that have been in orbit since the 1950s and remained in orbit as of June 2024. This unique study used TLE data spanning three solar cycles—Solar Cycle 22, Solar Cycle 23, and Solar Cycle 24—to examine the impact of solar activity on their orbital decay. The results highlight the key role of solar-driven thermospheric variability in determining orbital lifetimes. A consistent threshold in solar activity—when sunspot numbers exceed approximately 67–75\% of their cycle peak—was found to mark the onset of significantly accelerated orbital decay. This critical solar input level offers valuable insights for improving debris re-entry forecasting and satellite mission planning. The observed decline in decay rates from Cycle 22 to Cycle 24 closely mirrors the corresponding decrease in solar activity, as indicated by sunspot numbers and F10.7 flux values, highlighting the sensitivity of orbital dynamics to long-term solar variability. Additionally, the need to apply scaling factors to the ballistic coefficients to match the modeled and observed decay profiles points to limitations in both the historical TLE-based estimation methods and the MSIS 2.0 atmospheric model. While the model demonstrates reasonable accuracy for low- to mid-inclination orbits, significant discrepancies observed in high-inclination cases suggest that it inadequately represents thermospheric conditions in polar regions. Overall, this study underscores the complex interplay between solar activity, atmospheric density, and orbital decay, and highlights the need for improved atmospheric modeling to enhance prediction accuracy—particularly for objects in polar and high-latitude orbits.

\section*{Acknowledgments}
The authors express their sincere gratitude to the University of Colorado Space Weather Technology, Research, and Education Center (SWx TREC) for providing the Python wrapper for the MSIS 2.0 model.

\bibliographystyle{jasr-model5-names}
\biboptions{authoryear}
\bibliography{refs}

\end{document}